\documentclass[11pt]{article}
\usepackage{amsmath,amssymb}
\usepackage[dvips]{epsfig}
\usepackage[dvips]{color}
\usepackage{pb-diagram,lamsarrow,pb-lams}
\DeclareFontFamily{OT1}{rsfs}{}
\DeclareFontShape{OT1}{rsfs}{m}{n}{ <-7> rsfs5 <7-10> rsfs7 <10-> rsfs10}{}
\DeclareMathAlphabet\mathcurl{OT1}{rsfs}{m}{n}
\def\1#1{{\bf #1}}
\def\2#1{{\cal #1}}\def\9#1{{\sl #1}}\def\4#1{{\tt #1}}\def\5#1{{\sf #1}}
\def\6#1{{\mathfrak #1}}\def\7#1{{\mathbb #1}}\def\8#1{{\rm #1}}
\def\9#1{{\mathcurl #1}}

\def\3{{\ss}}

\def\ol{\overline}

\def\skb{\vskip 0.5cm}

\def\beq{\begin{eqnarray}}
\def\eeq{\end{eqnarray}}
\def\vs{\vspace{0.2cm} \\}

\newtheorem{The}{Theorem}[section]
\newtheorem{Def}[The]{Definiton}
\newtheorem{Lem}[The]{Lemma}
\newtheorem{Pro}[The]{Proposition}
\newtheorem{Cor}[The]{Corollary}
\def\bdef{\begin{Def}\1: \em}
\def\eef{\end{Def}}

\def\blem{\begin{Lem}\1: }
\def\elem{\end{Lem}}
\def\bthe{\begin{The}\1: }
\def\ethe{\end{The}}
\def\bpro{\begin{Pro}\1: }
\def\epro{\end{Pro}}
\def\bcor{\begin{Cor}\1: }
\def\ecor{\end{Cor}}

\def\al{\alpha}
\def\be{\beta}
\def\Gam{\Gamma}
\def\lam{\lambda}\def\Lam{\Lambda}
 
\def\te{\theta}

\def\sgm{\sigma}
\def\om{\omega}\def\Om{\Omega}

\def\bpr{\paragraph*{\it Proof.}}

\def\epr{$\square$\skb}

\def\pa{\partial}
\def\<{\langle}
\def\>{\rangle}
\def\Ad{{\rm Ad}}

\def\bdes{\begin{enumerate}}
\def\edes{\end{enumerate}}
\newcommand\itno[1]{\item[{\it ({#1})}]}

\def\bmat{\left( \begin{array}{ccc} }
\def\emat{\end{array} \right)}
\def\bcase{\left\{ \begin{array}{ccc} }
\def\ecase{\end{array} \right\}}

\def\beqa{\begin{eqnarray*}}
\def\eeqa{\end{eqnarray*}}
\def\bdia{\begin{diagram}}
\def\edia{\end{diagram}}

\def\ul{\underline}

\def\rMapsto{ \ \longmapsto \ }
 
\title{\bf Euclidean field theory on a sphere}
\author{{\it Dirk Schlingemann} \\
The Erwin Schr\"odinger International Institute \\ 
for Mathematical Physics (ESI)\\
Vienna}
\begin{document}
\maketitle
\abstract{ 
This paper is concerned with a structural analysis 
of euclidean field theories on the euclidean sphere.
In the first section we give proposal for axioms for a euclidean 
field theory on a sphere in terms of C*-algebras.
 
Then, in the second section, we investigate the 
short-distance behavior of euclidean field theory models
on the sphere by making use of the concept of {\em scaling 
algebras}, which has first been introduced by D. Buchholz, and 
R. Verch and which has also be applied to euclidean field theory
models on flat euclidean space in a previous paper. 
We establish the expected statement that
that scaling limit theories of euclidean field theories
on a sphere are euclidean field theories
on flat euclidean space.
 
Keeping in mind that the minkowskian analogue of the euclidean 
sphere is the de Sitter space, we develop a 
Osterwalder-Schrader type construction scheme
which assigns to a given euclidean field theory on the sphere
a quantum field theory on de Sitter space. 
We show that the constructed quantum field theoretical 
data fulfills the so called geodesic KMS condition 
in the sense of H. J. Borchers and D. Buchholz, i.e.
for any geodesic observer the system looks like 
a system within a thermal equilibrium state.}

\newpage
\section{Introduction}
\label{sc1}
One basic motivation for studying structural aspects of
euclidean field theory models on a sphere 
is that the finite volume of the sphere can be regarded as 
a natural infra-red regularizator. In fact, there are 
indications that non-trivial 
euclidean field theory models with an infra-red cutoff can be constructed. 
This is based on the work of J. Magnen, V. Rivasseau, and 
R. S\'en\'eor \cite{MagRivSen93}
where it is claimed that the Yang-Mills$_4$ model exists within a 
finite euclidean volume.

\paragraph{\em Euclidean field theory on the sphere.}
In order to give an overview of the ideas and strategies we 
use, a brief description of the setup, we are going to use, is given 
here. Euclidean field theory on a sphere can be formulated 
in a analogous manner as euclidean field theory on $\7R^d$
\cite{Schl97}. For a precise formulation of the axioms, we refer the reader
to Section \ref{sc2}. 
The mathematical ingredients which model the concepts 
of euclidean field theory consist of two main 
objects, a C*-algebra and a particular class 
of states on it.

The C*-algebra $\6B$, the first ingredient, 
has the following structure:
To each region $\9V$, for our purpose $\9V$
is a subset of the $d$-dimensional euclidean sphere,
a C*-subalgebra $\6B(\9V)\subset\6B$ is assigned.
We require that this assignment
is isotonous and local in the sense that, $\9V_1\subset\9V_2$ implies
$\6B(\9V_1)\subset\6B(\9V_2)$ 
and operators which are localized in disjoint regions commute, i.e.
$\9V_1\cap\9V_2=\emptyset$ implies $[\6B(\9V_1),\6B(\9V_2)]=\{0\}$.
In addition to that, the  assignment
$\9V\mapsto\6B(\9V)$ has a symmetry. The rotation group 
$\8O(d+1)$ acts covariantly on the algebra $\6B$ by 
automorphisms $\be_h$, here $h$ is a rotation, such that 
the algebra $\6B(\9V)$ is mapped via $\be_h$ onto the algebra of 
the corresponding rotated region $\6B(h\9V)$.

The states, which are of interest for our considerations, 
are {\em rotation invariant reflexion positive regular states} $\eta$.
Rotation invariance just means that $\eta$ is invariant 
under the automorphisms $\be_h$. The properties  
reflexion positivity and regularity are precisely formulated
in Section \ref{sc2}. Roughly speaking, both   
reflexion positivity and regularity impliy 
cretain analytic properties within 
variables upon which particular correlation functions depend. 
This fact is an essential ingredient for constructing 
a quantum field theory from euclidean data.

\paragraph{\em Scaling algebras and renormalization group.}
A general approach for the analysis of the high energy properties
of a given quantum field theory model has been developed by 
D. Buchholz and R. Verch \cite{Bu94,BuVer95,Bu96b,Bu96a,BuVer97,Bu97}. 
Such a short distance analysis can analogously be carried out
for euclidean field theory models \cite{Schl99}.

We show in
Section \ref{sc3} 
that the scaling limit theories of euclidean field theories
within a finite volume, here on a sphere, are 
essentially independent of the volume cutoff,
the radius of the sphere. As a result 
euclidean field theories on flat euclidean space $\7R^d$ 
are the result of the scaling limit procedure. 
This is what one expects, and we mention at this point that 
the analogous situation has already been studied 
for the analogous situation in Minkowski space \cite{BuHePr}. 
More precisely, the scaling limits of a 
quantum field theory in de Sitter spacetime are 
quantum field theories in flat Minkowski space. 

For a point $e\in rS^d$, the stabilizer subgroup of $e$ in 
$\8O(d+1)$ is isomorphic to $\8O(d)$.  If the scaling limit procedure 
is performed at $e$, the invariance under the stabilizer 
subgroup should remain as a  $\8O(d)$ invariance within the scaling 
limit. The translation invariance should then enter from the 
fact that the state under consideration $\eta$ 
is invariant under the full group $\8O(d+1)$.
The scaling limit procedure which we going to use 
can also be seen as blowing up the radius of the sphere
and we always choose $r=1$ for the unscaled theory. 

\paragraph{\em Quantum field theory on de Sitter spacetime.}
Keeping in mind that the minkowskian analogue of the euclidean 
$d$-sphere $S^d\subset\7R^{d+1}$ is the de Sitter space, 
we show in Section \ref{sc4}, 
by exploring the analytic structure of de Sitter space, 
that from a given euclidean field  
$(\6B,\be,\eta)$ on the sphere $S^d$ 
a quantum field theory $(\6A,\al,\om)$
on de Sitter space can be constructed.
The constructed state $\om$ satisfies the 
{\em geodesic KMS condition} which means that for any geodesic 
observer the state $\om$ looks like an equilibrium state.
These type of states have been analyzed by H. J. Borchers 
and D. Buchholz \cite{BorBu98}.  
A constructive example which fits perfectly within our axiomatic
framework has been given by R. Figari, R. H\"oegh-Krohn, and C. R. Nappi
\cite{FigHoehNap75}.
\section{Formulation of the axioms}
\label{sc2}
The starting point in the framework of algebraic
euclidean field theory is an isotonous net 
\beqa
\9V\rMapsto \6B(\9V)\subset \6B
\eeqa
of C*-subalgebras of $\6B$, indexed by convex  sets
$\9V\subset S^d$. This net covers the 
kinematical aspects of a particular model.
We require the following properties for the net:

\paragraph{\em $\8O(d+1)$-covariance:}
There exists a group homomorphism $\be $ from the 
orthogonal group $\8O(d+1)$ into the automorphism group of 
$\6B$ such that for each convex ste $\9V\subset S^d$ one has 
\beqa
\be_h\6B(\9V)&=&\6B(h\9V)
\eeqa
for each $h\in\8O(d+1)$. 

\paragraph{\em Locality:}
If $\9V_1\cap\9V_2=\emptyset$, then $[\6B(\9V_1),\6B(\9V_2)]=\{0\}$. 
\skb

We consider a particular class of states on $\6B$ which 
contain the dynamical information of the particular 
model. In order to formulate the 
analogue property of reflexion positivity in the euclidean approach in 
flat space, we consider 
for each point $e\in S^d$ on the sphere the half space 
\beqa
\8H_e^{d+1}&:=&\7R_+e+e^\perp\subset\7R^{d+1}
\eeqa
in the ambient space and we build the half sphere  
\beqa
\8{HS}^d_e&:=&\8H_e^{d+1}\cap S^d \ \ .
\eeqa
We denote by $\6B(e)$ the C*-algebra generated by operators
which are localized in $\8{HS}^d_e$. The reflexion 
$\te_e\in \8O(d+1)$ at the hyperplane 
$P^d_e=\8H_e^{d+1}\cap \8H_{-e}^{d+1}$ maps 
$\8{HS}^d_e$ onto $\8{HS}^d_{-e}$ and it leaves the 
hypersphere $S^{d-1}_e=\8{HS}^d_e\cap \8{HS}^d_{-e}$ stable.

We are now prepared to formulate sufficient  properties, shared by 
those states on $\6B$, which enable us to construct  
quantum field theory model from the euclidean data. 

\paragraph{\em Rotation invariance:}
A state $\eta$ on $\6B$ is called euclidean invariant
if for each $h\in \8O(d+1)$ the identity 
$\eta\circ\be_h=\eta$ holds true.

\paragraph{\em Reflexion positivity:}
A state $\eta$ on $\6B$ is called reflexion positive if  
exists a point $e\in S^d$ on the sphere   
such that the sesquilinear form
\beqa
\6B(e)\otimes\6B(e)\ni b_0\otimes b_1\mapsto 
\<\eta,j_e(b_0)b_1\>
\eeqa
is positive semi definite. Here, $j_e$ is 
the anti-linear involution which is given by
$j_e(b)=\be_{\te_e}(b^*)$.

\paragraph{\em Regularity:}
A state $\eta$ on $\6B$ is called regular if for each 
$b_0,b_1,b_2\in\6B$ the map
\beqa
h&\mapsto&<\eta,b_0\be_h(b_1)b_2>
\eeqa
is continuous. 
\skb

A triple $(\6B,\be,\eta)$ consisting of a euclidean net of 
C*-algebras $(\6B,\be)$ and a euclidean invariant reflexion positive
regular state $\eta$ is called a {\em euclidean field}
on the sphere $S^d$.

\section{Short-distance analysis for EFTh on $S^d$}
\label{sc3}
For $e\in S^d$ we consider the coordinate chart 
\beqa
\phi_e:\8{HS}^d_e\ni x &\mapsto& x-(ex)e\in\7R^d
\eeqa
where $P_e^d$ is identified with $\7R^d$ in a canonical manner.
For an element $h\in\8O(d+1)$ we get
\beqa
h\circ\phi_e&=&\phi_{he}\circ h \ \ .
\eeqa
In particular, $e$ is mapped to $0$ via $\phi_e$.
For $e_1\in S^d$, $e_1 e =0$, the hypersphere $S^{d-1}_{e_1}$
contains $e$. As a consequence 
$S^{d-1}_{e_1}\cap \8{HS}^d_e$ is mapped into the 
hyperplane $P^{d-1}_{\phi_e(e_1)}\subset\7R^d$
and  $\8{HS}^{d-1}_{e_1}\cap \8{HS}^d_e$ is mapped into the 
halfspace $\8H^{d-1}_{\phi_e(e_1)}\subset\7R^d$.

From a given euclidean field $(\6B,\be,\eta)$ on the sphere $S^d$
and a point $e\in S^d$, we obtain 
a net $\9U\mapsto \6B_e(\9U)$ of C*-algebras in a natural manner
by setting  
\beqa
\6B_e(\9U)&:=&\6B(\phi_e^{-1}(\9U))
\eeqa
for each bounded convex set $\9U\subset\7R^d$, 
where we use the convention $\6B(\emptyset):=\7C\11$.
Let $\be_e$ be the restriction of $\be$ to the 
stabilizer subgroup $\8O_e(d)\cong\8O(d)$ of $e$, 
then it is obvious that 
\beqa
\be_{(e,h)}\6B_e(\9U)&=&\6B_e(h\9U)
\eeqa
is valid for each $h\in\8O(d)$. Moreover, we have in general
\beqa
\be_h\6B_e(\9U)&=&\6B(h\phi_e^{-1}(\9U))\ \ = \ \ 
\6B_e(\phi_e h\phi_e^{-1}\9U)
\eeqa
for each $h\in\8O(d+1)$ and for each bounded convex set $\9U\subset\7R^d$ with
$h\phi_e^{-1}(\9U)\subset \8{HS}^d_e$.
The restriction $\eta_e=\eta|_{\6B_e}$ of $\eta$ to the 
C*-subalgebra $\6B_e$ is a reflexion positive $\8O(d)$-invariant regular
state.

\paragraph{\em Limit functionals.}
A convenient method for labeling the different scaling limit theories
makes use of {\em limit functionals}\cite{Schl99} (for the limit 
$\lam\to 0$ in $\7R_+$).
These functionals are states $\zeta$ on the C*-algebra of $\9F_\8b(\7R_+)$
of all bounded functions on $\7R_+$, which
annihilate the closed ideal $\9F_0(\7R_+)$,
which is generated by functions $f\in\9C_\8b(\7R_+)$ 
with $\lim_{\lam\to 0}f(\lam)=0$. Indeed, for a function $f\in\9F_\8b(\7R_+)$
with $\lim_{\lam\to 0}f(\lam)=f_0$, we find 
$<\zeta,f>=f_0$ for each limit functional $\zeta$. 
Since each limit functional can be regarded
as a measure on the spectrum of $\9F_\8b(\7R_+)$, we write 
\beqa
<\zeta,f>&=&\int\8d\zeta(\lam) \ f(\lam)
\eeqa
in a suggestive manner. 

\paragraph{\em Taking scaling limits.}
We briefly review here, how scaling limit models can be constructed
from the data $(\ul{\6B}_e,\be_e,\eta_e)$. 
First, we consider the C*-algebra 
of bounded $\6B$-valued functions on $\7R_+$, $\9F_\8b(\7R_+,\6B)$.
We introduce for a bounded convex set 
$\9U\subset\7R^d$ by $\ul{\6B}_e(\9U)$ the C*-subalgebra in  
$\9F_\8b(\7R_+,\6B)$ which is generated by functions 
\beqa
\lam\mapsto \1b(\lam)&=&\int \8dh \ f(h) \ \be_h\1b_o(\lam)
\eeqa
such that $\1b(\lam)\in\6B_e(\lam\9U)$ for each $\lam$. 
Here $f\in\9C^\infty(\8O(d+1))$ is a smooth function on $\8O(d+1)$ and 
$\8d h$ is the Haar measure on $\8O(d+1)$.
The C*-algebra which is generated by all 
local algebras $\ul{\6B}_e(\9U)$ is $\ul{\6B}_e$.
For a limit functional  
$\zeta$, we introduce the ideal $\6J_\zeta$
in $\ul{\6B}_e$ which consists of those functions
$\1b$ for which the C*-seminorm
\beqa
\|\1b\|_\zeta&=&\int\8d\zeta(\lam) \ \|\1b(\lam)\|
 \eeqa
vanishes. The scaling algebra $\6B_{(e,\zeta)}$ is just given by the 
quotient $\ul{\6B}_e/\6J_\zeta$ and $\1p_\zeta$
denotes in the subsequent the 
corresponding canonical projection onto the quotient.
We formally interpret $\6B_{(e,\zeta)}$
in terms of a direct integral decomposition with respect to the measure 
$\zeta$ and we write 
 \beqa
\1p_\zeta[\1b]&=&\int^\oplus \8d\zeta(\lam) \ \1b(\lam) \ \ .
\eeqa
The local algebras are given by 
$\6B_{(e,\zeta)}(\9U):=\1p_\zeta[\ul{\6B}_e(\9U)]$. 
The group homomorphism $\be^o_{(e,\zeta)}$ is given according to 
\beqa
\be^o_{(e,\zeta,h)}\1p_\zeta[\1b]
&=&\int^\oplus \8d\zeta(\lam) \ \be_h\1b(\lam)
\eeqa
for each $h$ in the stabilizer subgroup $\8O_e(d)$.
According to \cite{Schl99}, there exists a $\8O(d)$-invariant
reflexion positive state $\eta_{(e,\zeta)}$ 
on  $\6B_{(e,\zeta)}$ which is uniquely determined by 
\beqa
\<\eta_{(e,\zeta)},\1p_\zeta[\1b]\>&=&
\int \8d\zeta(\lam) \ \<\eta_e,\1b(\lam)\>
\eeqa
for each $\1b\in\ul{\6B}_e$. We are now prepared to formulate the 
following statement:

\bthe\label{the0}
There exists a group homomorphism $\be_{(e,\zeta)}$ 
from the euclidean group $\8E(d)$ into the automorphism group of
$\6B_{(e,\zeta)}$ which acts covariantly on the net 
$\ul{\6B}_{(e,\zeta)}$ and which extends the homomorphism 
$\be^o_{(e,\zeta)}$ such that the triple
$(\ul{\6B}_{(e,\zeta)},\be_{(e,\zeta)},\eta_{(e,\zeta)})$
is a euclidean field on $\7R^d$.
\ethe

\paragraph{\em Sketch of the proof.}
We postpone the complete proof of the theorem to Appendix \ref{app0}.
We briefly sketch here the main idea of the proof which is quite simple.
In order to construct an action of the translation group 
in $\7R^d$ we make use of the rotations which do not 
leave the point $e$ stable. We choose a orthonormal basis  
$(e_0,\cdots,e_d)$ with $e_0=e$. 
Let $L_{0\mu}$ be the generator of the rotations in the plane 
spanned by $e_0,e_\mu$ then for each $\mu=1,\cdots,d$ an automorphism 
on $\6B_{(e,\zeta)}$ is given by 
\beqa
\be_{(e,\zeta,se_\mu)}\1p_\zeta[\1b]
&=&\int^\oplus \8d\zeta(\lam) \ \be_{\exp(\lam s L_{0\mu})}\1b(\lam)
\eeqa
for each $s\in\7R_+$. Indeed, it turns out that these automorphisms generate 
an action of the translation group. 
In addition to that it can be shown that 
the automorphisms $\be^o_{(e,\zeta,h)}$, where $h$ is in the stabilizer
subgroup of $e$, together with the automorphisms 
$\be_{(e,\zeta,se_\mu)}$, $\mu=1,\cdots,d$, generate 
an action of the full euclidean group $\8E(d)$ on $\6B_{(e,\zeta)}$.

This is exactly what one expects by looking at the 
geometrical situation. Taking the scaling limit at the point $e$
can also be interpreted as blowing up the sphere $S^d$.
Heuristically, the spheres $\lam^{-1}S^d$ tend to 
$\7R^d$ if the radius $\lam^{-1}$ becomes infinite, i.e. $\lam\to 0$, 
and the point $e$ is identified with the origin $x=0$ in $\7R^d$.
During this limit process, 
the stabilizer subgroup of $e$ becomes the rotation group in $\7R^d$
and the remaining rotations, generated by $L_{0\mu}$, $\mu=1,\cdots,d$,
can be identified with the translations in $\7R^d$.

\section{From EFTh on the sphere to QFTh on de Sitter space}
\label{sc4}
This section is devoted to an analogous 
construction procedure as in \cite{Schl97} which 
relates a given euclidean field $(\6B,\al,\eta)$ on the sphere to a
quantum field theory in de Sitter spacetime.

According to our axioms, the map 
\beqa
\6B(e)\otimes\6B(e)\ni b_1\otimes b_2\mapsto \<\eta,j_e(b_1)b_2\>
\eeqa
is a positive semidefinite sesquilinear form.  
By dividing the null-space and taking the closure 
we obtain a Hilbert space $\2H$. The corresponding 
canonical projection onto the quotient is denoted by 
\beqa
\Psi:\6B(e)\mapsto \2H 
\eeqa
and we write $\Om:=\Psi[\11]$.
The construction 
of the observables, which turn out to be bounded operators on $\2H$, 
can be performed in several steps.

\paragraph{\em Construction of a representation of $\8{SO}(d,1)$.} 
The construction of a unitary strongly continuous representation  
$U$ of $\8{SO}(d,1)$ can be 
performed by applying the 
the theory of virtual group representations, as it has been
worked out by J. Fr\"ohlich, K. Osterwalder, and 
E. Seiler \cite{FrohOstSeil}, to our situation.  
This leads to the result:

\bthe\label{the00}
There exists a unitary strongly continuous representation 
$U$ of the Lorentz group $\8{SO}(d,1)$ on $\2H$.
\ethe
\bpr
A strongly continuous unitary 
representation $W$ of the stabilizer subgroup $\8O_e(d)\subset \8O(d,1)$
of $e$ can easily be constructed according to 
\beqa
W(h)\Psi[b]&=&\Psi[\be_h b] 
\eeqa
where $b$ is an operator in $\6B(e)$. 
In order to construct the Lorentz boosts, 
we introduce the regions $\Gam(e,\tau)$, $\tau\in (0,\pi/2)$, 
which is the intersection $\hat\Gam(e,\tau)\cap S^d$, where 
$\hat\Gam(e,\tau)$ is the $\8O(d+1)$ invariant 
cone in $e$ direction with opening angle $2\tau$. 
We choose an orthonormal basis 
$(e_0,\cdots, e_d)$ with $e=e_0$ and writing $L_{0\mu}$ for the 
generator of the rotations in the plane spanned by $e_0,e_\mu$, 
on obtains a vector valued function 
\beqa
\Psi_{(b,\mu)}(\8i s)&=&\Psi[\be_{\exp(sL_{0\mu})}b]
\eeqa
which is defined for each $b\in \6B(\Gam(e,\tau))$ and for
$|s|<\tau$. The function $\Psi_{(b,\mu)}$ has an holomorphic extension 
into the strip $\7R+\8i(-\tau,\tau)$.

Assuming that the net $\9V\mapsto\6B(\9V)$ fulfills 
{\em weak additivity} in the sense that for each convex set $\9V\subset S^d$
we have 
\beqa
\6B&=&\ol{\bigcup_{h\in\8O(d+1)}\6B(h\9V)}^{\|\cdot\|} \ \ , 
\eeqa
then, as we show in the Appendix \ref{app1}, the space 
$\2D(e,\tau):=\Psi[\6B(\Gam(e,\tau))]$ is dense in $\2H$
which allows to apply the results, shown in \cite{FrohOstSeil},
directly to our case.
In fact, as it has been carried out in \cite{FrohOstSeil}, 
the analytic properties of one parameter groups of boosts
can be exploit to get the result:
A unitary strongly continuous representation $U$ of 
the Loretz group $\8{SO}(d,1)$ is uniquely determined by
\beqa
U(h)&=&W(h)
\vs\vs
U(\exp(tB_\mu))\Psi[b]&=&\Psi_{(b,\mu)}(t)
\eeqa
where $h$ is an element of $\8{SO}_e(d)$ and $B_\mu$ is the generator 
of the Lorentz boosts leaving the  
wedge $\{x\in\7R^{d+1}||x^0|\leq x^\mu\}$ invariant.
\epr

\paragraph{\em Construction of the local net of observables.} 
In order to keep our technical assumptions as simple as possible 
we assume that the euclidean net $(\6B,\be)$ fulfills the 
time zero condition (TZ). This condition states that 
the C*-algebra is generated by the $j_e$-invariant elements 
$b\in\6B(S^{d-1}_e)$,
which are contained in the intersection $\6B(e)\cap\6B(-e)$, 
together with the transformed operators $\be_h b$, $h\in\8O(d+1)$.
The algebra $\6B(S^{d-1}_e)$ is represented by bounded operators 
on $\2H$ where the representation $\pi$ is given as follows:
\beqa
\pi(b)\Psi[b_1]&=&\Psi_\eta[bb_1] \ \ .
\eeqa
Analogously to the situation in Minkowski spacetime \cite{Schl97}
we assign to each bounded causally complete region 
$\9O\subset\8{dS}^d$ in $d$-dimensional de Sitter spacetime 
$\8{dS}^d$ the von Neumann algebra $\6A(\9O)\subset\6B(\2H)$
which is generated by the 
bounded operators $U(g)\pi(b)U(g)^*$
where $b$ is localized in  a convex set $\9G$ of the 
time slice $S^{d-1}_e$ and $g\9G\subset\9O$.
The C*-algebra which is generated by all local algebras 
$\6A(\9O)$ ($\9O\subset\8{dS}^d$ causally complete and bounded) 
is denoted by $\6A$.  

We also obtain a group homomorphism $\al$ form 
the Lorentz group $\8{SO}(d,1)$ into the automorphism group 
$\8{Aut}\6A$ by setting $\al_g:=\8{Ad}U(g)$ for each 
Lorentz transformation $g$.  
By construction  $\al$ acts covariantly on the net
$\9O\mapsto\6A(\9O)$, i.e. for each 
causally complete region $\9O$ in de Sitter 
space, the automorphism $\al_g$ maps the algebra 
$\6A(\9O)$ onto $\6A(g\9O)$.   

\paragraph{\em The geodesic KMS condition.}
There is a canonical Lorentz invariant state $\om$ on $\6A$
which is just given by $\<\om,a\>:=\<\Om,a\Om\>$.
For a Boost generator $B_\mu$ we denote by 
$\9W_{B_\mu}:=\{x\in\8{dS}^d||x^0|\leq x^\mu\}$ the intersection of 
the de Sitter space $\8{dS}^d$ with the wedge in ambient space  
associated to the boosts $\exp(sB_\mu)$.
We also consider for a boost generator $B=B_\mu$ and the corresponding
one-parameter group $\al_B$ of automorphisms 
\beqa
\al_{(B,t)}&:=& \al_{\exp(tB)} 
\eeqa
which obviously maps the algebra $\6A(\9W_B)$ onto itself and
we get a W*-dynamical system $(\6A(\9W_B),\al_B)$. 
In the subsequent we prove that  $\om$ fulfills 
the  geodesic KMS condition which can be precisely formulated by the theorem
which is proven in Appendix \ref{app2}:

\bthe\label{the1}
The restricted state $\om|_{\6A(\9W_B)}$ 
is a KMS state with respect to the W*-dynamical system $(\6A(\9W_B),\al_B)$
at inverse temperature $2\pi$.
\ethe

\paragraph{\em The modular conjugation associated with a wedge algebra.}
The geodesic KMS condition (Theorem \ref{the1}) implies that 
the vector $\Om$ is cyclic and separating for 
the wedge algebra $\6A(\9W_B)$ (compare \cite{Schl99b}). We build the 
modular conjugation $J_B$ as well as the modular operator
$\Delta_B$ with respect to the pair $(\6A(\9W_B),\Om)$.
We choose an orthonormal basis $(e_0,\cdots,e_d)$ with $e=e_0$.

For $B=B_\mu$ the intersection 
$\8{HS}^{d-1}_{(e_0,e_\mu)}:=S^{d-1}_{e_0}\cap\8{HS}^d_{e_\mu}$ 
is the spatial base of the wedge $\9W_B$. 
The reflexion $\te_{e_\mu}$ at the hyperplane
$e_\mu^\perp$ is contained in the stabilizer group 
of $e=e_0$ and the prescription 
\beqa
\9J_B\Psi[b]&=&\Psi[j_{e_\mu}(b)]
\eeqa
defines a anti-unitary operator $\9J_B$, a PCT operator, on $\2H$. 
Following the analysis, carried out in \cite{Schl99b}, 
one finds (see Appendix \ref{app3}:

\bthe\label{the2}
The modular conjugation $J_B$ of the pair $(\6A(\9W_B),\Om)$
coincides with the PCT operator $\9J_B$:
\beqa
J_B&=&\9J_B \ \ .
\eeqa
\ethe

\paragraph{\em Verification of the Haag-Kastler axioms.}
The statement of Theorem \ref{the2} can be used
to verify the Haag-Kastler axioms for the 
net $\9O\mapsto\6A(\9O)$ in a very straight forward 
manner. We already know that there exists a 
group homomorphism $\al$ form the Lorentz group $\8{SO}(d,1)$ into the 
automorphism group of $\6A$ which is covariant with respect to the net 
structure, i.e. $\al_g\6A(\9O)=\6A(g\9O)$ is valid 
for each causally complete set $\9O$ in de Sitter space and for each 
$g\in\8{SO}(d,1)$. 
It remains to be proven that locality is satisfied which is formulated in the 
corollary:

\bcor\label{cor1}
The net $\9O\mapsto\6A(\9O)$ fulfills locality, i.e.
if $\9O_1$ and $\9O_2$ are spacelike separated regions in 
de Sitter space, then the commutator $[a_1,a_2]=0$ vanishes 
for each $a_1\in\6A(\9O_1)$ and for each $a_2\in\6A(\9O_2)$.
\ecor
\bpr
The geometric action of the PCT operator 
$\9J_B$ implies that 
\beqa
\9J_B\6A(\9W_B)\9J_B&=&\6A(\9W_B')
\eeqa
where $\9W_B'$ is the causal complement of $\9W_B$ in de Sitter space.
Since $\9J_B$ coincides with the modular conjugation $J_B$ by
Theorem \ref{the2}, we conclude
\beq\label{wd}
\6A(\9W_B')&=&\6A(\9W_B)' \  \ .
\eeq
The net  $\9O\mapsto\6A(\9O)$ is $\8{SO}(d,1)$ covariant and thus  
Equation (\ref{wd}) holds true for each wedge region 
$\9W$ in de Sitter space. 
Choosing $\9W$ in such a way that 
$\9O_1\subset \9W$ and $\9O_2\in\9W'$, it follows that $[a_1,a_2]=0$  
for each $a_1\in\6A(\9O_1)\subset\6A(\9W)$ 
and for each $a_2\in\6A(\9O_2)\subset\6A(\9W')=\6A(\9W)'$.
\epr

\section{Conclusion and outlook}
\label{sc5}

\paragraph{\em Conclusion:}
We have proposed to consider a finite volume euclidean 
field theory in $d$ dimensions as a 
field theory on the $d$-sphere $S^d\subset\7R^{d+1}$.
The corresponding euclidean net $\6B$ 
carries a covariant action of the $d+1$-dimensional rotation group 
$\8O(d+1)$ and the functional $\eta$ is invariant under this action. 
For a given point $e\in S^d$ on the sphere and 
for a given limit functional $\zeta$, we have constructed the 
scaling limit theory
$(\6B_{(e,\zeta)},\be_{(e,\zeta)},\eta_{(e,\zeta)})$
at $e$. The invariance under the stabilizer 
subgroup of $e$ remains as a  $\8O(d)$ invariance of the state
$\eta_{(e,\zeta)}$ whereas the  
translation invariance enter from the 
fact that the underlying state $\eta$ 
is invariant under the full group $\8O(d+1)$.
This leads to the result, we expected, namely that 
the 
scaling limit theory of a euclidean field theory on the sphere is 
a euclidean field theory in an infinite volume.
 
Moreover, we have discussed how to construct a 
quantum field theory $(\6A,\al,\om)$
in de Sitter space form a given euclidean field theory 
$(\6B,\be,\eta)$ on the sphere,  
by exploring the analytic structure of de Sitter space. 
In particular, we have proven that the reconstructed state $\om$
fulfills the so called {\em geodesic KMS condition}, i.e.
for any geodesic observer the state $\om$ looks like 
an equilibrium state.

\paragraph{\em Outlook:}
Alternatively, one can consider a euclidean field theory within a 
compact region $\Lam\subset\7R^d$ with boundary $\pa\Lam$.
The corresponding euclidean net of C*-algebras 
$\9U\mapsto\6B_\Lam(\9U)$ 
is then indexed by convex regions $\9U\subset\7R^d$.
By choosing the region $\Lam$ rotationally invariant, it makes sense
to consider an action of the rotation group $\8O(d)$ 
by automorphisms $\be_{(\Lam,h)}$ on the algebra $\6B_\Lam$.
The axiom of $\8O(d)$- invariance and reflexion positivity 
can analogously by formulated for a state $\eta_\Lam$.

For a given limit point $\zeta$, the corresponding 
scaling limit theory
$(\6B_{(\Lam,\zeta)},\be_{(\Lam,\zeta)},\eta_{(\Lam,\zeta)})$
at $x=0\in\Lam$ can be constructed. There are two 
natural questions which one can ask within this context:
\bdes

\itno 1
Is the scaling limit theory
$(\6B_{(\Lam,\zeta)},\be_{(\Lam,\zeta)},\eta_{(\Lam,\zeta)})$
a euclidean field theory on $\7R^d$, i.e. within an infinite volume,
where $ \eta_{(\Lam,\zeta)}$ is invariant under the full euclidean 
group $\8E(d)$?

\itno 2
Do the scaling limit theory
$(\6B_{(\Lam,\zeta)},\be_{(\Lam,\zeta)},\eta_{(\Lam,\zeta)})$
depend on the choice of boundary conditions at $\pa\Lam$?
\edes

\subsubsection*{{\em Acknowledgment:}}
I am grateful to Prof. Jakob Yngvason for 
supporting this investigation with hints and many ideas.
I would also like to thank Prof. Detlev Buchholz for 
useful hints and remarks. 
This investigation is financially supported by the 
{\em Jubil\"aumsfonds der Oesterreichischen Nationalbank} 
which is also gratefully acknowledged.
Finally I would like to thank the 
Erwin Schr\"odinger International Institute for Mathematical Physics, 
Vienna (ESI) for its hospitality.
\newpage
\begin{appendix}
\section{On a Reeh-Schlieder-type theorem for local euclidean algebras.}
\label{app1}
Regularity, reflexion positivity, and the euclidean invariance of the 
functional $\eta$ imply certain analytic properties of correlation 
functions. This leads to a Reeh-Schlieder-type theorem 
for local euclidean algebras, by 
assuming that the net $\9V\mapsto\6B(\9V)$ fulfills 
{\em weak additivity} in the sense that for each 
convex set $\9V\subset S^d$:
\beqa
\6B&=&\ol{\bigcup_{h\in\8O(d+1)}\6B(h\9V)}^{\|\cdot\|} \ \ .
\eeqa

\bthe\label{rs}
For each bounded open convex set $\9V\subset\8{HS}^d_e$ 
contained in half sphere $\8{HS}^d_e$
the subspace  $\2D(\9V):=\Psi[\6B(\9V)]$ is dense in $\2H$.
\ethe 
\bpr
The proof can be obtained by an application of a 
Reeh-Schlieder-type argument. 
Consider a unit vector $e_1$, perpendicular to 
$e$. 
Then we define linear operators 
\beqa
V_{e_1}(s):\2D(\9V)&\to &\2H
\eeqa
according to 
\beqa
V_{e_1}(s)\Psi[b]&=&\Psi[\be_{\exp(sL_{ee_1})}b] 
\eeqa
for $s\in I(e_1,\9V)$, where the open interval $I(\9V)$ is given by
\beqa
I(e_1,\9V)&:=&\{s|\exp(sL_{ee_1})\9V\subset\8{HS}_e^d\}
\eeqa
and $L_{ee_1}$ denotes the generator of rotations in the 
plane spanned by $e,e_1$.
One easily checks that $V_{e_1}(s)$ is symmetric, i.e.
\beqa
\<\Psi_1,V_{e_1}(s)\Psi_2\>&=&\<V_{e_1}(s)\Psi_1,\Psi_2\>
\eeqa
for each $\Psi_1,\Psi_2\in\2D(\9V)$  and we get
\beqa
s-\lim_{s\to 0;s\in I(e_1,\9V)}V_{e_1}(s_1)\Psi&=&\Psi
\eeqa
for each $\Psi\in\2D(\9V)$.
Due to a theorem by J. Fr\"ohlich \cite{Froh80} or 
by using the results of A. Klein and L. J. Landau \cite{KlLan81a}, 
the operators $V_{e_1}(s)$ extends uniquely 
to self adjoint operators on $\2H$. 
This implies that the vector valued function 
\beqa
\8is\mapsto V_{e_1}(s)\Psi 
\eeqa
has an holomorphic extension into the open strip $\7R+\8i I(\9V)$
for each $\Psi\in\2D(\9V)$.
More general, for operators $b_1,\cdots,b_k\in\6B(\9V)$, 
the operator valued function 
\beqa
\Psi_{[b_1,\cdots,b_k; e_1,\cdots,e_k]}:\8i(s_1,\cdots,s_k)\mapsto 
\Psi\biggl[\prod_{j=1}^k\be_{\exp(s_kL_{ee_k})}b_k\biggr]
\eeqa
has an holomorphic extension into the tube
$\7R^{k}+\8iI(e_1,\cdots,e_k,\9V)$, where 
the region $I(e_1,\cdots,e_k,\9V)$ contains
all points $(s_1,\cdots,s_k)\in\7R^k$ such that 
\beqa
\exp(s_1L_{ee_1})\9V(s_2,e_2;\cdots;s_k,e_k)\subset \8{HS}_e^d \ \ .
\eeqa
The set $\9V(s_2,e_2;\cdots;s_k,e_k)$ is recursively defined by
\beqa
\9V(s_2,e_2;\cdots;s_k,e_k)&:=&
\exp(s_2L_{ee_2})[\9V\cup \9V(s_3,e_3;\cdots;s_k,e_k)] \ \ .
\eeqa
In particular, by construction 
$I(e_1,\cdots,e_k,\9V)$ is an open connected set.

Let $\Psi'$ be a vector in the orthogonal 
complement of $\2D(\hat\9V)$, where $\hat\9V$ is a slightly larger 
region than $\9V$. Then we conclude that there is an open connected subset 
$J\subset I(e_1,\cdots,e_k,\9V)$ such that
\beqa
\<\Psi',\Psi_{[b_1,\cdots,b_k; e_1,\cdots,e_k]}(z)\>&=&0
\eeqa
for each $z\in\7R^k+\8iJ$. Since $\Psi_{[b_1,\cdots,b_k; e_1,\cdots,e_k]}$
is holomorphic in the tube $\7R^{k}+\8iI(e_1,\cdots,e_k,\9V)$
we conclude 
\beqa
\<\Psi',\Psi_{[b_1,\cdots,b_k; e_1,\cdots,e_k]}(z)\>&=&0
\eeqa
for all $z\in\7R^{k}+\8iI(e_1,\cdots,e_k,\9V)$.  
By making use of the weak additivity of the net, the 
set of vectors
\beqa
\biggl\{\Psi_{[b_1,\cdots,b_k; e_1,\cdots,e_k]}(\8is)
\biggm|b_1,\cdots,b_k\in\6B(\9V);
s\in I(e_1,\cdots,e_k,\9V),k\in\7N\biggr\}
\eeqa
span a dense subspace in $\2H$ which implies $\Psi'=0$
and the theorem follows.
We mention here that a similar argument can also be found in
\cite{KlLan81}. 
\epr

\section{Proof of Theorem \ref{the0}}
\label{app0}
We consider an orthonormal basis of $\7R^{d+1}$, $(e_0,\cdots,e_d)$, $e=e_0$.
Let $\be_{(e_i,e_j)}:s\mapsto\be_{(e_i,e_j,s)}$, be the 
one-parameter group of automorphisms, related to the  
rotations in the $e_i-e_j$-plane.
The corresponding generators in the Lie algebra $\6o(d+1)$ are 
denoted by $L_{ij}$. 
For a scaling parameter $\lam\in\7R_+$ and for a bounded convex
region $\9U\subset\7R^d$ we consider the set $G(\lam,\9U)\subset\7R$ 
which consists of all $s\in\7R$ such that 
$\exp(\lam s L_{0\mu})\phi_e^{-1}(\lam\9U)\in\8{HS}_e^d$.
It is obvious that the definition is independent of the choice  
of the coordinate direction $\mu=1,\cdots , d$ and that 
the inclusion $G(\lam,\9U)\subset G(\lam',\9U)$ is valid 
for $\lam'<\lam$.
For a function $\1b\in\ul{\6B}_e(\9U)$ we put
\beqa
[\ul{\be}_{(e,se_\mu)}\1b](\lam)
&:=& \bcase \be_{\exp(\lam s L_{0\mu})}\1b(\lam) &;& s\in G(\lam,\9U) \\
                   \1b(\lam) &;& s\not\in G(\lam,\9U) \ecase 
\eeqa
which defines a function in $\9F_\8b(\7R_+,\6B)$.
In order to verify that $\ul{\be}_{(e,se_\mu)}\1b$ is contained 
in $\ul{\6B}_e$, we compute for $x\in\9U$ and for each 
$s\in G(\lam,\9U)$
\beq\label{eq1}
&&[\phi_e\exp(\lam s L_{0\mu})\phi_e^{-1}(\lam x)]^\nu
\vs\nonumber\vs
&=&\bcase
\lam x^\nu &;& \mu\not=\nu \\
\lam \cos(\lam s)x^\mu + \sin(\lam s)(1-\lam^2x^2)^{1/2}&;& \mu=\nu
\ecase
\nonumber\vs\nonumber\vs\nonumber
&=&\lam\cdot \bcase
x^\nu &;& \mu\not=\nu \\
\cos(\lam s)x^\mu + \lam^{-1}\sin(\lam s)(1-\lam^2x^2)^{1/2}&;& \mu=\nu
\ecase \ \ .
\eeq
We introduce the region 
\beqa
\tau_{se_\mu}\9U&=&\bigcup_{\lam:s\in G(\lam,\9U)}
\lam^{-1}\phi_e\exp(\lam s L_{0\mu})\phi_e^{-1}(\lam \9U)
\eeqa
which is compact since $\lam^{-1}\sin(\lam s)(1-\lam^2x^2)^{1/2}=\8O(1)$.
This implies that $\ul{\be}_{(e,se_\mu)}\1b$
is contained in $\ul{\6B}_e(\tau_{se_\mu}\9U)$ and 
the prescription 
\beqa
\be_{(e,\zeta,se_\mu)}\1p_\zeta[\1b]
&=&\1p_\zeta[\ul{\be}_{(e,se_\mu)}\1b]
\eeqa
yields a well defined automorphism of $\6B_{(e,\zeta)}$.
Analogously we define 
an automorphism 
$\be_{(e,\zeta, s_1 e_{\mu_1}+\cdots +s_k e_{\mu_k})}$ by replacing 
$sL_{0\mu}$ by $s_1 L_{0\mu_1}+\cdots +s_k L_{0\mu_k}$
and we show that 
\beqa
\be_{(e,\zeta,s_1 e_\mu)}\al_{(e,\zeta,s_2e_\nu)}
&=&\be_{(e,\zeta,s_1e_\mu+s_2e_\nu)} 
\eeqa
which implies that for $x=\sum_\mu x^\mu e_\mu$ the assignment 
\beqa
x\mapsto \be_{(e,\zeta, x)}
\eeqa
is indeed a group homomorphism form the translation group $\7R^d$ 
into the automorphism group of $\6B_{(e,\zeta)}$.
Consider the function 
\beqa
\1b:\lam\mapsto \1b(\lam)&=&\int \8dh \ f(h) \ \al_h\1b_o(\lam)
\eeqa
then we get the estimate 
\beqa
&&\|\ul{\be}_{(e,s_1e_\mu)}\ul{\be}_{(e,s_2e_\nu)}\1b(\lam)
-\ul{\be}_{(e,s_1e_\mu+s_2e_\nu)}\1b(\lam)\|
\vs\vs
&\leq&\sup_{\lam\in\7R_+}\|\1b_o(\lam)\| 
\vs
&\times&
\int \8dh \ 
\biggm|
f(h)-f( \8e^{-\lam s_1 L_{0\mu}}\8e^{-\lam s_2 L_{0\nu}}
\8e^{\lam[s_1 L_{0\mu}+s_2 L_{0\nu}]}h)\biggm|  \ \ .
\eeqa
Let $V$ be a finite dimensional linear space on which 
$\8O(d+1)$ is represented by unitary operators. 
We may assume that the function $f$ is given by 
$f(h)=f_V(hv)$ with $v\in V$  and $f_V\in\9C^\infty_0(V)$. 
A straight forward computation shows that there exists 
a linear operator $M(\lam)\in L(V)$ with 
$\|M(\lam)\|\leq\8{const.}\lam^2$ such that 
\beqa
\8e^{-\lam s_1 L_{0\mu}}\8e^{-\lam s_2 L_{0\nu}}
\8e^{\lam[s_1 L_{0\mu}+s_2 L_{0\nu}]}h v
&=& hv+M(\lam)hv \ \ .
\eeqa
Hence we conclude 
\beqa
\lim_{\lam\to 0}
\sup_{h\in\8O(d+1)}\biggm|
f(h)-f( \8e^{-\lam s_1 L_{0\mu}}\8e^{-\lam s_2 L_{0\nu}}
\8e^{\lam[s_1 L_{0\mu}+s_2 L_{0\nu}]}h)\biggm|  &=& 0
\eeqa
which implies the desired result
\beqa
\1p_\zeta[\ul{\be}_{(e,s_1e_\mu)}\ul{\be}_{(e,s_2e_\nu)}\1b(\lam)
-\ul{\be}_{(e,s_1e_\mu+s_2e_\nu)}\1b(\lam)]&=&0 \ \ .
\eeqa
For an element $h$ of the stabilizer subgroup $\8O_e(d)$
one easily checks the relation
\beqa
\be^o_{(e,\zeta,h)}\be_{(e,\zeta,x)}&=&\be_{(e,\zeta, hx)}\be^o_{(e,\zeta,h)}
\eeqa
and the existence of the homomorphism $\be_{(e,\zeta)}$ follows.
It remains to be proven that $\be_{(e,\zeta)}$ acts covariantly 
on $\6B_{(e,\zeta)}$. 
Let a bounded convex set $\9U\subset\7R^d$ be given.
According to Equation (\ref{eq1}) we conclude that 
there exists $r>0$ such that 
\beqa
\lam^{-1}\phi_e\exp(\lam s L_{0\mu})\phi_e^{-1}(\lam \9U)
\subset (\9U+se_\mu)+B_d(r\lam)
\eeqa
for each $\lam\in\7R_+$. Here $B_d(r\lam)$ is the closed 
ball in $\7R^d$ with center $x=0$ and radius $r\lam$.
This implies 
\beqa
\be_{(e,\zeta,s e_\mu)}\6B_{(e,\zeta)}(\9U)&=&\6B_{(e,\zeta)}(\9U+se_\mu)
\eeqa
which proves the covariance.
Since the state $\eta_{(e,\eta)}$ is translationally invariant, 
which is due to the construction of $\be_{(e,\zeta)}$, the theorem follows.
\epr

\section{Proof of Theorem \ref{the1}}
\label{app2}
The main steps of the proof can be performed in complete analogy to the 
the analysis of \cite{Schl99b,KlLan81}. 
We consider a family of operators $b_1,\cdots, b_n$ which are contained 
in the time slice algebra, where 
$b_j\in\6B(\9G_j)$ is localized 
in a convex subset $\9G_j\subset S^{d-1}_e\cap\9W_{B_{\mu_j}}$. 
Let $s\mapsto \be_{(j,s)}$ be the one-parameter 
automorphism group, corresponding to the rotations in $\mu_j-0$ direction. 
This implies that 
$\be_{(j,s)}b_j\in\6B(\8{HS}^d_e)$ for each $s\in(0,\pi)$.
Let $B$ be the boost in $0-\mu_0$ direction, let 
$L$ be the generator of the rotations in the 
$0-\mu_0$ plane, and let $L_j$ be the 
generator in the of the rotations in the 
$0-\mu_j$ plane. We introduce the open subset in $\7R^2$
\beqa
I(\9G_j)&:=&\{(\tau,s)\in\7R^2|\forall j:
\exp(sL)\exp(\tau L_j)\9V_j\subset\8{HS}^d_e\}
\eeqa
which contains in particular the set $\{0\}\times(0,\pi)\subset I(\9G_j)$.
By introducing the operators 
\beqa
\1b_j(\tau_j)&:=&V_{\mu_j}(\tau_j)\pi(b_j)V_{\mu_j}(-\tau_j)
\eeqa
we obtain by an analogous computation as it has been 
carried out in \cite{Schl99b}:
\beqa
&&\<V(s_n)\1b_n(\tau_n)\cdots V(s_{k+1})\1b_{k+1}(\tau_{k+1})\Om,
V(s_k)\1b_k(\tau_k)\cdots
\vs\vs
&&\cdots V(s_1)\1b_1(\tau_1)\Om\>
\vs\vs
&=&
\<V(\pi-(s_1+\cdots +s_k))\1b_1(\tau_1)^*V(s_1)\1b_2(\tau_2)^*\cdots 
V(s_{k-1})\1b_k(\tau_k)^*
\Om,
\vs\vs
&&\times \ 
V(\pi-(s_{k+1}+\cdots +s_n))\1b_{k+1}(\tau_{k+1})^*
V(s_{k+1})\1b_{k+2}(\tau_{k+2})^*\cdots 
\vs\vs
&&\cdots V(s_{n-1})\1b_n(\tau_n)^*\Om\>
\eeqa
which expresses the KMS condition at inverse temperature $2\pi$ 
in the euclidean points.
Finally, a straight forward application of the analysis of \cite{KlLan81}
proves the theorem.
\epr

\section{Proof of Theorem \ref{the2}}
\label{app3}
By following the analysis of \cite{KlLan81}, we choose 
a family of operators $b_1,\cdots, b_n$ 
which are contained 
in the time slice algebra, where 
$b_j\in\6B(\9G_j)$ is localized 
in a convex subset $\9G_j\subset S^{d-1}_e\cap\9W_{B_{e_j}}$,
and we choose 
directions 
$e_0,\cdots, e_n$ which are perpendicular to $e$.
By using the same notations as for the proof of Theorem \ref{the1},
we obtain by putting $\9J:=\9J_B$:
\beqa
&&V(s_k)\1b_k(\tau_k)\cdots V(s_1)\1b_1(\tau_1)\Om
\vs\vs
&=&
V(s_k)\1b_k(\tau_k)V(-s_k)V(s_k+s_{k-1})\1b_{k-1}(\tau_{k-1})
V(-s_k-s_{k-1})\cdots
\vs\vs
&&\cdots V(s_1)\1b_1(\tau_1)\Om
\vs\vs
&=&
V(s_k)\1b_k(\tau_k)V(-s_k)\cdots V(s_1+\cdots+s_k)\1b_1(\tau_1)
\Om
\vs\vs
&=&
\Psi[b_k(\tau_k,s_k)\cdots b_1(\tau_1,s_1+\cdots+s_k)] \ \ .
\eeqa
We compute for $s_1,\cdots,s_k\in\7R_+$ and
$(\tau_i,s_k+\cdots+s_i)\in I(\9G_i)$ for $k\leq i\leq 1$:
\beqa
&&\9J V(s_k)\1b_k(\tau_k)\cdots V(s_1)\1b_1(\tau_1)\Om
\vs\vs
&=&
\Psi[j_{e_0}(b_k(\tau_k,s_k)\cdots b_1(\tau_1,s_1+\cdots+s_k))]
\vs\vs
&=&
\Psi
[b_1(\sgm_1\tau_1,\pi-(s_1+\cdots+s_k))^*\cdots b_k(\sgm_k\tau_k,\pi-s_k)^*]
\vs\vs
&=&
V(\pi-s_1+\cdots+s_k)\1b_1^*(\sgm_1\tau_1)V(s_1)
\cdots \1b_k^*(\sgm_k\tau_k)^*V(s_k)\Om 
\eeqa
with $\sgm_j=1$ if $e_j\perp e_0$ and $\sgm_j=-1$ if $e_j=e$.
Performing an analytic continuation within the parameter 
$s_1,\cdots s_k$ and $\tau_1,\cdots,\tau_k$ and taking
boundary values at $s_j=\tau_j=0$ yields the relation
(compare \cite{KlLan81} as well as \cite{FrohOstSeil} and \cite{Schl97}) 
\beqa
&&\9J_B
\biggl[\prod_{j=1}^k U(\exp(t_jB_{e_j}))\1b_jU(\exp(-t_jB_{e_j}))\biggr]
\Om
\vs\vs
&=&
V(\pi)
\biggl[\prod_{j=1}^k U(\exp(t_jB_{e_j}))\1b_jU(\exp(-t_jB_{e_j}))\biggr]^*
\Om 
\eeqa
which implies that the Tomita operator is 
\beqa
J_B\Delta_B^{1/2}&=&\9J_B V(\pi)
\ \ .
\eeqa
Moreover, according to Theorem \ref{the1},
the automorphism group 
\beqa
\al_B:t\mapsto \Ad[U(\exp(tB))] 
\eeqa
maps $\6A(\9W_B)$ into itself and the state $\om|_{\6A(\9W_B)}$
is a KMS state at inverse temperature
$2\pi$ and the theorem follows. 
\epr

\end{appendix}
\newpage


\end{document}